\documentclass{iaus}
\usepackage{graphicx}

\title[Can pulsational instabilities impact a massive star's rotational evolution?] %% give here short title %%
      {Can pulsational instabilities impact a massive star's rotational evolution?}

\author[Rich Townsend \& Jim MacDonald]   %% give here short author list %%
       {Rich Townsend$^{1,2}$ \and Jim MacDonald$^{2}$}

\affiliation{$^1$Bartol Research Institute, University of Delaware, Newark, DE 19716, USA\\
             email : {\tt rhdt@bartol.udel.edu} \\[\affilskip]
             $^2$Department of Physics \& Astronomy, University of Delaware, Newark, DE 19716, USA\\
             email : {\tt jimmacd@udel.edu}}

\pubyear{2008}
\volume{250}  %% insert here IAU Symposium No.
\pagerange{1--6}
\jname{Massive Stars as Cosmic Engines}
\editors{F. Bresolin, P. Crowther \& J. Puls, eds.}
\begin{document}

\maketitle

\begin{abstract}
We investigate whether angular momentum transport due to unstable
pulsation modes can play a significant role in the rotational
evolution of massive stars. We find that these modes can redistribute
appreciable angular momentum, and moreover trigger shear-instability
mixing in the molecular weight gradient zone adjacent to stellar
cores, with significant evolutionary impact.

\end{abstract}

\firstsection % if your document starts with a section,
              % remove some space above using this command..5

\section{Background: Pulsation in Massive Stars} \label{sec:intro}

As the many, extensive variability surveys of the past couple of
decades have revealed, pulsation in massive stars appears to be
ubiquitous.  Examples of these surveys include the \textit{HIPPARCOS}
astrometry mission, which photometrically discovered over a hundred
new pulsating B stars (e.g., \cite[Waelkens et~al. 1998]{Wae1998});
the study by \cite[Fullerton et~al. (1996)]{Ful1996}, revealing
optical line-profile variations consistent with pulsation in 23 out of
a sample of 30 O stars; and the \emph{IUE} Mega campaign (\cite[Massa
  et~al. 1995]{Mas1995}), which highlighted systematic variability in
the wind of the early B supergiant HD~64760, subsequently attributed
to co-rotating interaction regions rooted in photospheric pulsations
(\cite[Fullerton et~al. 1997]{Ful1997}; see also \cite[Kaufer
  et~al. 2006]{Kau2006}).

Against this observational background, it seems reasonable to
conjecture that those O and B stars already confirmed as pulsators
could represent just the tip of the iceberg --- that, in fact, a far
greater proportion of massive stars are undergoing pulsations, albeit
at amplitudes that fall below present-day detection thresholds.  This
expectation is lent considerable support by theoretical calculations
(e.g., \cite[Pamyatnykh 1999]{Pam1999}, his Figs. 3 \&~4) showing that
any star with a mass $M_{\ast} \gtrsim 3\,M_{\odot}$ \emph{must} pass
through one or more pulsation instability strips as it evolves from
ZAMS to TAMS.

These instability strips all arise from the operation of a
thermodynamic engine within the star, which converts radiant heat into
mechanical energy associated periodic pulsation. As \cite[Eddington
  (1926)]{Edd1926} originally pointed out (`\textit{\ldots we require,
  in fact, something corresponding to the valve-mechanism of a heat
  engine\ldots}'), a key component of this engine is a regulatory
process that adds heat to the stellar material when at its hottest,
and removes heat when at its coolest. In classical ($\delta$) Cepheid
pulsators, the regulatory process is the positive temperature
dependence of the Rosseland mean opacity $\kappa$ at temperatures
$\log T \approx 4.5$ where second helium ionization occurs. For
massive pulsators, a similar `$\kappa$ mechanism' operates on the
opacity peak at $\log T \approx 5.3$ associated with bound-bound
transitions of iron-group elements. This `iron bump' leads to
overstable p-mode pulsations in the $\beta$ Cepheid stars ($M_{\ast}
\gtrsim 7\,M_{\odot}$; \cite[Dziembowski \& Pamyatnykh
  1993]{DziPam1993}), and to g-mode pulsations in the slowly pulsating
B (SPB) stars ($3\,M_{\odot} \lesssim M_{\ast} \lesssim 7\,M_{\odot}$;
\cite[Dziembwoski et~al. 1993]{Dzi1993}) and in supergiant B stars
($M_{\ast} \gtrsim 25\,M_{\odot}$; \cite[Pamyatnykh
  1999]{Pam1999}). Here, the quoted mass ranges are for modes of
harmonic degree $\ell = 0\ldots2$; toward larger values of $\ell$, the
SPB and supergiant g-mode instability strips merge (see \cite[Balona
  \& Dziembowski 1999]{BalDzi1999}).

\section{Wave Transport of Angular Momentum}

Traditionally, massive-star pulsation has been regarded simply as a
dynamical phenomenon to be modeled: we see variations in the
photospheric or wind diagnostics of a particular star, and we attempt
to interpret these variations as arising from pulsation
perturbations. More recently, the advent of specialized space
observatories such as \textit{MOST} (\cite[Walker
  et~al. 2003]{Wal2003}) and \textit{COROT} (\cite[Baglin
  et~al. 2006]{Bag2006}) has opened the door to applying the
techniques of asteroseismology to massive stars --- using the
oscillation spectrum of a pulsating star to place constraints on
interior physics such as the incidence of convective overshoot, or the
degree of differential rotation.

In both of these contexts, pulsation is seen as a passive player in a
star's evolution. But what if, conversely, the star's evolutionary
trajectory were determined to some extent by its pulsation?  This idea
has already been applied to low-mass stars; \cite[Talon \& Charbonnel
  (2003, 2005)]{TalCha2003,TalCha2005}, for instance, argue that
internal gravity waves (IGWs --- essentially, g-mode transients damped
over a timescale commensurate with their period) play a role in
braking the rotation in the inner regions of such stars.

To include the effects of IGWs on stellar evolution, an extra term is
added to the equation governing angular momentum transport, so that
\begin{equation} \label{eqn:transport}
\rho \frac{{\rm d}}{{\rm d} t} \left[ r^{2} \Omega \right] =
\frac{1}{5 r^{2}} \frac{\partial}{\partial r} \left[ \rho r^{4} \Omega
  U \right] +
\frac{1}{r^{2}} \frac{\partial}{\partial r} \left[ \rho \nu r^{4}
\frac{\partial \Omega}{\partial r} \right] - 
\frac{3}{8\pi} \frac{1}{r^{2}} \frac{\partial}{\partial r}
\mathcal{L}_{J}.
\end{equation}
This equation applies to shellular differential rotation (as argued by
\cite[Zahn 1992]{Zah1992}, strong horizontal turbulence will tend to
enforce uniform rotation across shells of constant radius $r$).
Modulo geometrical factors of order unity, the term on the left-hand
side represents the local rate of change of angular momentum per unit
radius, with $\Omega(r)$ the local angular velocity. On the right-hand
side, the first term represents angular momentum transport due to
meridional circulation with a velocity $U(r)$. The second term
represents diffusive processes with a transport coefficient $\nu$; the
major contribution to $\nu$ comes from convection and, in radiative
zones, from secular shear instability (e.g., \cite[Maeder \& Meynet
  1996]{MaeMed1996}). Finally, the third term represents wave
transport, as described by a luminosity function $\mathcal{L}_{J}(r)$
that quantifies the net amount of angular momentum carried per unit
time through the shell at radius $r$. The main contribution to
$\mathcal{L}_{J}$ comes from the Reynolds stress,
\begin{equation} \label{eqn:L_J}
L_{J} = 4 \pi r^{2} \rho \langle r \sin \theta\, \mathbf{v}_{r}\,
\mathbf{v}_{\phi} \rangle.
\end{equation}
Here, $\mathbf{v}_{r}$ and $\mathbf{v}_{\phi}$ are the radial and
azimuthal velocity perturbations due to the wave, and $\langle\rangle$
denotes the average over all solid angles.

\section{Application to Massive Stars}

In low-mass stars, stochastic processes such as turbulent stresses or
convective penetration are considered as the dominant wave excitation
mechanism (e.g., \cite[Talon \& Charbonnel 2003]{TalCha2003}).  Many
authors have assumed that the same processes (albeit operating in
different parts of the interior) are responsible for wave excitation
in massive stars; for instance, \cite[Maeder \& Meynet
  (2000)]{MaeMey2000} remark that `\textit{\ldots we could expect
  gravity waves to be generated by turbulent motions in the convective
  core.}'

However, as should be clear from \S\ref{sec:intro}, the waves observed
in massive stars are not stochastic IGWs but unstable global standing
oscillations, driven to large amplitudes by the iron-bump $\kappa$
mechanism. Thus, the appropriate formalism for treating wave transport
in these stars is a normal mode analysis with inclusion of excitation
and damping processes --- that is, nonradial, nonadiabatic
pulsation theory. This was clearly recognized by \cite[Ando (1986, and
  self-references therein)]{And1986}, who was the first to conjecture
that nonadiabatic pulsation may play an important role in shaping the
internal rotation profile of massive stars. Unfortunately, Ando's
investigations predated the release (in the early 1990's) of updated
opacity data that revealed the iron bump; thus, he was unable to reach
any firm conclusions.

To build on this prior work, we examine angular momentum transport by
high-order, intermediate-degree g modes in a $10\,M_{\odot}$ stellar
model near the end of its main-sequence evolution ($X_{\rm core} =
0.02$). (There is nothing particularly significant about this
$M_{\ast}$; our results generalize to stars of both higher and lower
masses. However, the late evolutionary stage is chosen to emphasize
the deposition of angular momentum in the molecular weight gradient
zone, discussed further below).
We use the \textsc{boojum} pulsation code (see \cite[Townsend
  2005]{Tow2005}) to calculate the complex oscillation spectrum of the
stellar model. Modes whose eigenfrequency $\omega$ has a negative
imaginary part (i.e., $\Im(\omega) < 0$) are unstable; the
eigenfunctions of these modes encapsulate all of the information
necessary to evaluate the $\mathbf{v}_{r}$ and $\mathbf{v}_{\phi}$
terms in eqn.~(\ref{eqn:L_J}), with the exception of an arbitrary
overall normalization.

\begin{figure}[t]
% \vspace*{-2.0 cm}
\begin{center}
\includegraphics{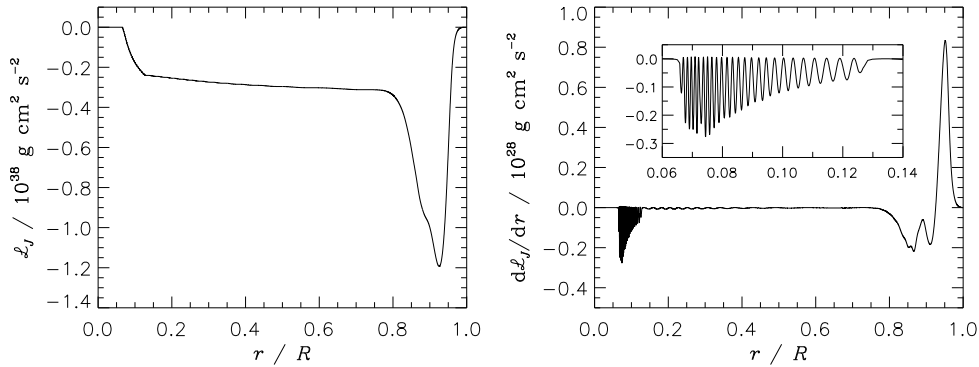} 
% \vspace*{-1.0 cm}
 \caption{The angular momentum luminosity $\mathcal{L}_{J}$ (left),
   and its radial derivative ${\rm d}\mathcal{L}_{J}/{\rm d}r$
   (right), plotted as a function of radius for the
   $\{n,\ell,m\}=\{40,4,-4\}$ g mode of the $10\,M_{\odot}$ model. The
   inset in the right-hand panel details the variation of the
   luminosity derivative in the $\mu$-gradient zone adjacent to the
   core.} \label{fig:luminosity}
\end{center}
\end{figure}

Fig.~\ref{fig:luminosity} plots the angular momentum luminosity
$\mathcal{L}_{J}$ as a function of fractional radius $r/R_{\ast}$ for
a single unstable g mode of the $10\,M_{\odot}$ model, having indices
$\{n,\ell,m\} = \{40,4,-4\}$ and normalized so that the peak
photospheric velocity perturbation is $1\,{\rm km\,s^{-1}}$ (this is a
conservative choice; for reference, the typical photospheric
velocities observed in pulsating massive stars are on the order of the
sound speed, $\sim 10-20\,{\rm km\,s^{-1}}$). Also plotted is the
luminosity derivative ${\rm d}\mathcal{L}_{J}/{\rm d}r$; as
eqn.~(\ref{eqn:transport}) indicates, this quantity is positive where
angular momentum is extracted, and negative where it is deposited. The
figure reveals angular momentum extraction from the surface layers,
where the $\kappa$ mechanism excites the g mode, and matching angular
momentum deposition in the interior, primarily in two regions where
the g mode is strongly damped. The outer damping region ($0.78
\lesssim r/R_{\ast} \gtrsim 0.92$) arises from the $\kappa$ mechanism
operating in reverse: the opacity has a strongly negative temperature
dependence, and so the thermodynamic engine converts mechanical energy
into radiant heat. The inner damping region ($0.07 \lesssim r/R_{\ast}
\lesssim 0.13$) is associated with the zone of varying molecular
weight ($\mu$) adjacent to the convective core. In this zone, the g
mode has a very short wavelength due to the steep gravitational
stratification; this leads to a spatially oscillatory pattern of
angular momentum deposition, as can be seen from the inset in the
right-hand panel of Fig.~\ref{fig:luminosity}.

The angular momentum luminosity shown in Fig.~\ref{fig:luminosity}
reaches a peak magnitude of $1.2 \times 10^{38}\,{\rm
  g\,cm^{2}\,s^{-2}}$; by way of comparison, \cite[Talon \& Charbonnel
  (2003, their Fig.~4)]{TalCha2003} find a net luminosity of $\sim 2
\times 10^{36}\,{\rm g\,cm^{2}\,s^{-2}}$ for IGWs in their model for a
$1.2\,M_{\odot}$ star. The two orders-of-magnitude difference between
these values is simply a reflection of the far-higher amplitudes
associated with the unstable modes found in massive stars, than the
stochastically excited waves in low-mass stars. To give a rough
estimate of the expected impact of the unstable modes, we note that
over $1\,{\rm Myr}$ (a typical timescale for main-sequence evolution)
the total angular momentum deposited in the $\mu$-gradient zone by the
$\{n,\ell,m\} = \{40,4,-4\}$ mode would, \textit{ceteris paribus}, be
on the order of $8 \times 10^{50}\,{\rm g\,cm^{2}\,s^{-1}}$. This is
approaching the total angular momentum $\sim 10^{51}\,{\rm
  g\,cm^{2}\,s^{-1}}$ stored in the core if the $10\,M_{\odot}$ star
were rotating uniformly at the critical rate. Thus, the angular
momentum transport due to the pulsation can be expected to have an
appreciable impact on the star's rotational evolution, and --- at the
most general level --- the answer to the question posed in this
paper's title is in the affirmative.

\section{A Self-Consistent Simulation}

\begin{figure}[t]
% \vspace*{-2.0 cm}
\begin{center}
\includegraphics{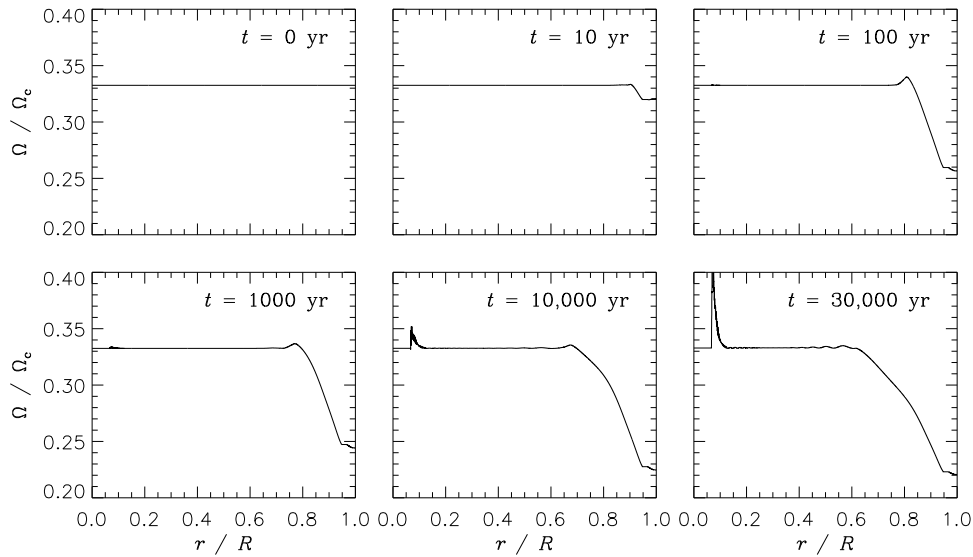} 
% \vspace*{-1.0 cm}
 \caption{Snapshots of the angular velocity $\Omega$ of the
   $10\,M_{\odot}$ model, plotted as a function of radius at six
   epochs during the \textsc{heimdall}
   simulation.} \label{fig:snapshots}
\end{center}
\end{figure}

Of course, the estimates given above neglect the fact that as the
internal rotation profile evolves in response to the angular momentum
transport, there will be a corresponding feedback effect on the
pulsation. Clearly, some kind of self-consistent simulation is
desirable, and to this end we have developed a prototype
pulsation-transport code. The code, named \textsc{heimdall}, solves
the angular momentum transport equation~(\ref{eqn:transport}) for an
input stellar model. The meridional circulation term is neglected,
because we are interested in transport occurring on timescales shorter
than the circulation timescale $R_{\ast}/U$; however, the diffusion term is
retained, to allow the rotation profile to relax from the steep
angular velocity gradients created by the wave transport term.  To
evaluate the wave transport term, a modularized version of the
\textsc{boojum} code is used to calculate the complex oscillation
spectrum of the stellar model at each simulation timestep. As in the
preceding section, the angular momentum luminosity is obtained from
mode eigenfunctions; however, rather than arbitrarily fixing mode
amplitudes, \textsc{heimdall} allows them to evolve over each timestep
in accordance with individual linear growth/damping rates
$-\Im(\omega)$.

\begin{figure}[t]
% \vspace*{-2.0 cm}
\begin{center}
\includegraphics{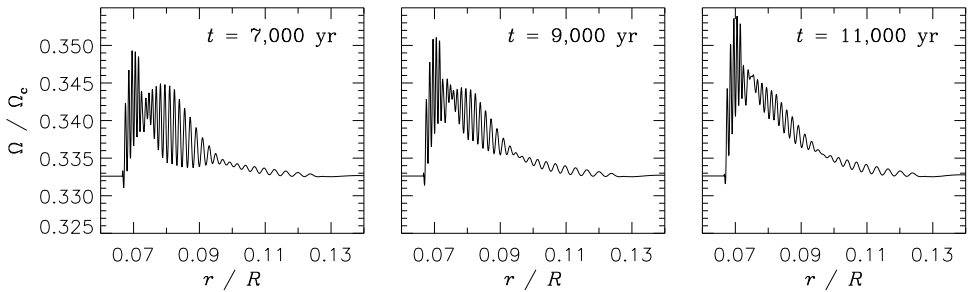}
% \vspace*{-1.0 cm}
 \caption{The evolution of the angular velocity $\Omega$ in the
   $\mu$-gradient zone, for the same simulation shown in
   Fig.~\ref{fig:snapshots}. Note how the steep shears in the
   left-hand panel have been mixed away by the shear instability in
   the center and right-hand panels.} \label{fig:mixing}
\end{center}
\end{figure}

Fig.~\ref{fig:snapshots} shows snapshots of the rotation profile
$\Omega(r)$ for the $10\,M_{\odot}$ model, from a \textsc{heimdall}
simulation of transport by $\{\ell,m\} = \{4,-4\}$ g modes. The
simulation begins in a state of uniform rotation at 33\% of the
critical rate $\Omega_{\rm c}$. Initially, a broad spectrum of g
modes, with radial orders $n = 26\ldots47$, are unstable toward the
$\kappa$ mechanism. As these g modes grow in amplitude, they transport
angular momentum inwards from the surface layers. Because these layers
contain little mass, they are braked quite rapidly; this established a
broad shear region separating the interior from the surface, which
acts to damp all but one of the initially unstable g modes. The radial
order of the single remaining mode progressively increases from $n =
47$ to $n = 63$ in a sequence of mode-switching episodes; in between
the switching, the mode hovers at the borderline of neutral stability,
maintaining a surface amplitude of $\sim 1-2\,{\rm km\,s^{-1}}$.

In the $10,000\,{\rm yr}$ panel of Fig.~\ref{fig:snapshots}, the
long-term impact of the single remaining mode begins to emerge:
angular momentum is deposited in the $\mu$-gradient zone, resulting in
its gradual spin-up. For the reasons discussed previously the
deposition is spatially oscillatory, and leads to the establishment of
nested shear layers of very narrow extent ($\sim 10^{-3}\,R_{\ast}$). These
shear layers are clearly revealed in the left-hand panel of
Fig.~\ref{fig:mixing}; however, in the center and right-hand panels,
the shear layers have been partly dissolved by diffusive transport
associated with the secular shear instability, which tends to smooth
out steep gradients in $\Omega$.

This result represents perhaps the most exciting finding in our
exploratory calculations. As it dissolves shear layers established by
pulsation angular momentum transport, the shear instability will mix
the chemical composition in the $\mu$-gradient zone. Given that this
zone plays a pivotal role in modulating angular momentum transport, in
particular serving as an insulator which inhibits meridional
circulation coupling between core and envelope, \emph{we expect that
  the disruption of this zone by shear/pulsation-assisted mixing
  (SPAM) will have a profound impact on the rotational evolution of
  massive stars}. Building on the foundation established by our
\textsc{heimdall} simulations, we plan further calculations to examine
the precise nature of this impact.

\vspace*{0.1cm}

\noindent \textbf{Acknowledgments:} RHDT is supported by NASA grant
          {\it LTSA}/NNG05GC36G.

% Bibliography

% Discussion

\begin{discussion}

\discuss{Cranmer}{Are your plots showing the equatorial plane? If so,
  might the angular momentum be circulating back toward the surface at
  mid-latitudes, say?}

\discuss{Townsend}{No, since the rotation is shellular, the angular
  velocity is uniform over each spherical surface.}

\discuss{Maeder}{[In considering angular momentum transport by g
    modes], if we account for horizontal turbulence, with the
  coefficient by S. Mathys and myself, it introduces a strong damping
  factor which considerably reduces the efficiency of this transport
  process.}

\discuss{Townsend}{I think that the horizontal turbulence will be
  efficient at damping IGWs excited stochastically in the core; but
  the g modes I'm considering are excited by the iron-bump $\kappa$
  mechanism in the envelope, and are far more robust. Don't forget
  that we see direct observational evidence for these unstable g
  modes.}

\discuss{Skinner}{Could you comment on what observational data are now
  available or might be available in the near future to test the
  validity of the models/simulations?}

\discuss{Townsend}{Survey data will be most useful in looking for
  evidence (e.g., a correlation between pulsation and surface
  enrichment) that the $\mu$-gradient zone has been disrupted by
  SPAM. In this respect, both the Large Synoptic Survey Telescope and
  the Kepler mission look to be promising developments.}

\end{discussion}

\end{document}